\documentclass{elsart}

\usepackage{graphicx,natbib,amssymb,lineno}

\makeatother

\begin{document}
\newcommand{\nvidiaNS}{NVIDIA}
\newcommand{\nvidia}{NVIDIA}
\newcommand{\geforce}{GeForce}

\begin{frontmatter}

\title{Thermodynamical properties of the Undulant Universe}
\author[1]{Tian Lan}
\address[1]{Department of Astronomy, Beijing Normal University, Beijing, 100875, P.R.China}
\author[1,2]{Tong-Jie Zhang\corauthref{cor1}},
\corauth[cor1]{Corresponding author. Tel.: +86 010 58805258; fax:
+86 010 58806319.}\ead{tjzhang@bnu.edu.cn}
\author[3]{Bao-Quan Wang}
\address[2]{Center for High Energy Physics, Peking University, Beijing 100871, P. R. China}
\address[3]{Department of Physics, Dezhou University, Dezhou, 253023, P. R. China}

\begin{abstract}
\setlength{\baselineskip}{1.6\baselineskip}Recent observations
show that our universe is accelerating by dark energy, so it is
important to investigate the thermodynamical properties of it. The
Undulant Universe is a model with equation of state
$\omega(a)=-\cos(b\ln a)$ for dark energy. Due to the term, $\ln
a$, neither the event horizon nor the particle horizon exists in
the Undulant Universe. However, the apparent horizon is a good
holographic screen, and can be seen as a boundary of keeping
thermodynamical properties. The Universe is in thermal equilibrium
inside the apparent horizon, i.e. the Unified First Law and the
Generalized Second Law of thermodynamics are confirmed. As a
thermodynamical whole, the evolution of the Undulant Universe
behaves very well in the current phase. However, considering the
unification theory and the real universe, the failure of
conversation law indicates that the model of the Undulant Universe
needs further consideration.

\end{abstract}


\end{frontmatter}

\section{Introduction}
\setlength{\baselineskip}{1.6\baselineskip} \label{intro} Many
cosmological observations \cite{1,2,3,4}, such as the type Ia
Supernova (SN Ia), Wilkinson Microwave Anisotropy Probe (WMAP),
the Sloan Digital Sky Survey (SDSS) etc. support that our Universe
is experiencing an accelerated expansion, and dark energy (DE)
with negative pressure, contributes about $72\%$ of the matter
content to the present Universe. For DE, we have the equation of
state (EOS) $P_{\Lambda} = \omega\rho_{\Lambda}$. However, the
nature of DE is known so little. In order to get more knowledge of
DE, many researchers have discussed it from thermodynamical
perspective widely. Such as thermodynamics of DE with constant
$\omega $ in the range $-1 < \omega < -1/3$ \cite{5}, $\omega =-1$
in the de Sitter spacetime and anti-de Sitter spacetime \cite {6},
$\omega <-1$ in the Phantom field \cite{7,8}, and $\omega =\omega
_{0}+\omega _{1}z$ \cite{9} and so on. More discussions on the
thermodynamics of DE can be found in \cite{10,11,12,13,14}. In
this paper, we also investigate a model from the thermodynamical
perspective to find some more interesting properties of DE. G. 't
Hooft found that the 3-dimensional world is an image of data. The
data can be stored on a 2-dimensional holographic film, and the
information on the projection is called hologram \cite{30}. Later,
Fischler and Susskind applied holography to the standard
cosmological context \cite{31}, and considered the world as a
hologram \cite{32}. The entire information about the global
3+1-dimensional spacetime can be stored on particular
hypersurfaces (called holographic screens). In other words, the
degrees of freedom of a spatial region reside on the surface of
the region. The total number of the degrees of freedom does not
exceed the Bekenstein-Hawking bound \cite{33}, which is a
universal entropy bound within a given weakly gravitational
volume. The holographic principle is helpful to understand the DE.

Usually we treat the Universe as a whole, and a global spacetime.
So some conclusions on BH can be used to the Universe. Hawking
temperature, BH entropy, and BH mass satisfy the First Law of
thermodynamics $dM=TdS$ \cite{21}. In terms of the definition of
the rationalized area \cite{22}, if a charged BH is rotating,
solving $dM$, we get $dM=(\kappa/8\pi)d\mathcal{A}+\Omega_{0}
dJ+V_{0}dQ$. Where $\kappa$, $\mathcal{A}$, $\Omega_{0}$, $J$,
$V_{0}$, $Q$ are the surface gravity, area, dragged velocity,
angular momentum, electric potential and charge of a BH
respectively. This expression is analogous with the First Law of
thermodynamics $dE = TdS-PdV$. This law suggests a connection
between thermodynamics and BH physics in general, and between
entropy and BH area in particular. Bekenstein conjectured these
analysis at first, and Hawking discovered that the BH can emit
particles according to the Planck spectrum. So we get the
effective temperature on the horizon of the BH, $T = \kappa/2\pi$
\cite{23}, and the entropy, $S = \mathcal{A}/4$ \cite{24}. Then we
can study many gravitational systems in the framework of
thermodynamics. Consequently, our Universe can be considered as a
thermodynamical system \cite{25,26,27,28}, and the thermodynamical
properties of BH can be generalized to spacetime enveloped by the
cosmological horizon. In other words, the thermodynamical laws
should be satisfied in the Universe.

We use the Planck units $c=G=k_{B}=\hbar=1$, where $G$ is Newton's
constant, $\hbar$ is Planck's constant, $c$ is the speed of light,
and $k_{B}$ is Boltzmann's constant respectively. The Planck units
of energy density, mass, temperature, and other quantities are
converted to CGS units.

This paper is organized as follows: In Sec. 2, we review the
introduction to the Undulant Universe. In Sec. 3, we discuss the
cosmological horizons of the Undulant Universe. In Sec. 4 and Sec.
5, we study the Unified First Law of thermodynamics and the
Generalized Second Law of thermodynamics on the apparent horizon
respectively. In Sec. 6, we present conclusions and discussions.

\section{Introduction to the Undulant Universe}
Our homogeneous and isotropic Universe follows the dynamics of an
expanding Robertson-Walker (RW) spacetime $(i,j=1,2)$:
\begin{eqnarray}
ds^{2} &=& g_{ij}dx^{i}dx^{j}+r^{*2}d\Omega^{2}\\
\nonumber &=& dt^{2}-a^{2}(t)(\frac{dr^{2}}{1-{\kappa}r^{2}}
+r^{2}d\theta^{2}+r^{2}\sin^{2}\theta d\phi^{2}). \label{rwm}
\end{eqnarray}Where we choose a class of comoving coordinate system
$(t,r,\theta,\phi)$ for RW metric, $x^{i}=(t,r)$ are the arbitrary
coordinates spanning the radial 2-spheres $(\theta,\phi) =
constant$, $a(t)$ is the expansion scale factor of the Universe,
$r^{*}=ar$, is defined in the usual way from the proper area of a
2-spheres $x^{i}=constant$: $\mathcal{A}=4\pi r^{*2}$, and
$d\Omega$ is the metric of 2-dimensional unit sphere respectively.
The spatial hypersurfaces of the Friedmann Universe have positive,
zero and negative curvatures for $\kappa= +1, 0$ and $-1$. And its
metric is:
\begin{equation}
g_{ij} = \left( \begin{array}{cc}
  -1 & 0  \\
  0  & \frac{a^{2}}{1-\kappa r^{2}}
\end{array}\right).
\label{h}
\end{equation}The evolution of the Universe is governed by the Friedmann equation:
\begin{equation}
H^{2}=\frac{\dot{a}^{2}}{a^{2}}=\frac{8{\pi}G}{3}\rho-\frac{\kappa}{a^{2}}.\label{fde}
\end{equation}Where $H$ is the Hubble parameter, the energy density $\rho$ is a
sum of different components: $\rho=\sum_{i}\rho_{i}$, evolving
differently as the Universe expands. To characterize the energy
density of each component, we define density parameter:
\begin{equation}
\Omega_{i}=\rho_{i}/\rho_{c},\label{dp}
\end{equation}where the critical density:
\begin{equation}
\rho_{c}=\frac{3H^{2}}{8\pi G}.\label{cd}
\end{equation} Observational evidence, extended by detailed
observations \cite{15}, presents a flat Universe. The mass-energy
of the Universe includes 0.05 ordinary matter and 0.22 nonbaryonic
dark matter, and is dominated by DE. The densities in matter and
vacuum are of the same order of magnitude. The phase of the
Universe transferred from radiation dominated to matter dominated
in the past, and recently, the vacuum energy is dominated at $z
\sim 1.5$. In other words, if the vacuum energy becomes dominated
at any other epoch, the Universe would have evolved completely
different history. To solve this cosmic coincidence problem, many
workers put forward many ideas \cite{16,17}, but all of them
involve fine tuning problem in some way. \cite{18}

To avoid these problems, we investigate Undulant Universe
characterized by alternating periods of acceleration and
deceleration \cite{19}, with an EOS:
\begin{equation}
\omega(a)=-\cos(b\ln a),\label{eos}
\end{equation}and the Hubble parameter:
\begin{eqnarray}
H^{2}(a)=H_{0}^{2}(\Omega_{M0}a^{-3}+\Omega_{\Lambda0}a^{-3}\exp[\frac{3}{b}\sin(b\ln
a)]).\label{h}
\end{eqnarray} Where $\Omega _{M0}$, $\Omega_{\Lambda0}$ are the density
parameter of matter and DE respectively, and the dimensionless
parameter $b$ controls the frequency of the accelerating epochs.
Note that in the limit of small values of $b$, the EOS above
approaches the cosmological constant, $\omega \sim -1$. As
discussed in \cite{20}, the best fit values for the parameters $b
= 0.06 \pm 0.01$, so we choose $b=0.05$ in the following
discussion. From FIG.1, we can see the evolution of $\omega$ for
the Undulant Universe. To demonstrate the nowadays Universe being
not particular clearly, we choose two periods, three periods, and
four periods respectively on large range of scales to discuss in
the following discussion.

At different epochs, $\omega$ behaves in the same way that
oscillates as a cosine function, and the average of $\omega$ over
one period is zero. Consequently, the cosmic coincidence problem
is solved, and no fine tuning is required. The range of possible
destinies for the Universe, even in the near case, is very broad
indeed. The Universe needn't evolve toward the cataclysm of
terminal inflation or recollapse necessarily, but might go on with
the sedate drift of a big slink.

\begin{figure}[t!]
\begin{center}
\includegraphics[width=8.5cm]{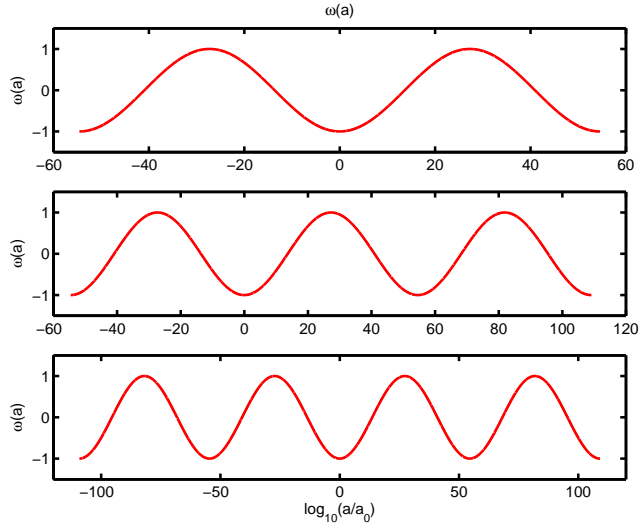}\\
\caption{Evolution of $\omega$ with $a$ in the Undulant Universe.
These three panels present it at different epochs, and they cover
two periods, three periods, four periods respectively.}
\label{fig1}
\end{center}
\end{figure}

\section{cosmological horizons of the Undulant Universe}
In general, global holographic screens are highly non-unique
\cite{35}. We concentrate on a class of Universes with
well-defined holographic screens of finite area. There are many
kinds of horizons for the boundary \cite{36}, and we will discuss
some of them below.

We concentrate on the quasi-local of the cosmological horizon.
Because the DE dominated Universe is asymptotic flatness at
infinity, in the asymptotic regions the matter is diluted.
Although there can be significant matter in the interior of the
spacetime, we can assume that there just DE contributes to the
energy on the quasi-local of the cosmological horizon, i.e.
$\Omega_{\Lambda0}=1$, $\Omega_{M0}=0$. So the Hubble parameter
becomes:
\begin{equation}
H^{2}(a)=H_{0}^{2}a^{-3}\exp[\frac{3}{b}\sin(b\ln a)].\label{ha}
\end{equation}There are two quite different horizon concepts in cosmology,
to which cosmologists have devoted their attention at various
times. Firstly, for a global observer, the radius of an event
horizon (EH) at cosmic time $t$ can be written as:
\begin{equation}
R_{E}=a(t)\int_{t}^{t_{f} }\frac{dt}{a(t)}.\label{re}
\end{equation}
When $\int_{t}^{t_{f}}dt/a(t) < \infty$. Where $t_{f}$ is the
final cosmic time, and generally, $t_{f}\rightarrow +\infty$.
Consequently, for the Undulant Universe:
\begin{equation}
\int_{t}^{t_{f}}\frac{dt}{a(t)}=\int_{a}^{+\infty}\frac{da}{a^{2}H}=\int_{a}^{+\infty}\frac{da}{H_{0}a^{0.5}\exp[\frac{1.5}{b}\sin(b\ln
a)]}, \label{rea}
\end{equation}Eq.(12) $\rightarrow +\infty$, which results from the oscillating
term, $\ln(a=\infty)\rightarrow +\infty$, so sine term has not a
certain value, and the integral is infinity. We can understand it
intuitively. Because we can't ensure the final phase of the
Universe, the radius is unknown. Different final phase can evolve
a different history, in other words, our recent Universe can
evolve into any phase in the future.
Secondly, the radius of a particle horizon (PH) at cosmic time $t$
can be written as:
\begin{equation}
R_{P}=a(t)\int_{t_{i}}^{t}\frac{dt}{a(t)}.\label{rp}
\end{equation}When $\int_{t_{i}}^{t}dt/a(t) < \infty$. Where $t_{i}$ is the
initial cosmic time, and generally, $t_{i}\rightarrow -\infty$.
For the Undulant Universe:
\begin{equation}
\int_{t_{i}}^{t}\frac{dt}{a(t)}=\int_{0}^{a}\frac{da}{a^{2}H}=\int_{0}^{a}\frac{da}{H_{0}a^{0.5}\exp[\frac{1.5}{b}\sin(b\ln
a)]}.\label{rpa}
\end{equation}Eq.(14) $\rightarrow \infty$. Similarly, the integral is
unknown because of the uncertainty of $\ln(a=0)$. We cannot give a
certain PH, i.e. any phase in the past can evolve into our recent
Universe. According to the discussion above, we find the Undulant
Universe really remove the coincidence problem, and the nowadays
phase is not particular any more. Whatever the phase is in the
past or in the future, our Universe can be the recent phase.

However, we have to find some other horizons to study the
thermodynamical properties of the Undulant Universe. Hawking and
Ellis define an apparent horizon (AH), which is the outer boundary
of a connected component of a trapped region. This 2-dimensional
AH can necessarily be extended to a 3-dimensional, time-evolved AH
over some finite range of $t$ \cite{46}. The AH is local in time,
and it is easy to locate in the numerical relativity. To simplify
the calculation, we define the physical radius as the form of
Eq.(3):
\begin{equation}
r^{*}=ar, \label{a4}
\end{equation}the hypersurface can be set as:
\begin{equation}\label{f}
f=g^{ij}r^{*}_{,i}r^{*}_{,j}.
\end{equation}For the AH, $f=0$. Then we get the radius of an AH is
\begin{equation}\label{ra}
r^{*}_{A}=ar_{A}=H_{0}^{-1}a^{1.5}\exp[-\frac{1.5}{b}\sin(b\ln
a)].
\end{equation}The radius depends on the details of the matter
distribution in the Universe. In generic situation, the AH evolves
with time and visibility of the outside antitrapped region depends
on the time development of the AH. Because there is no EH, the
spatial region outside of the horizon at a given time might be
observed. The change of the radius varying with time during a
Hubble time is:
\begin{equation}\label{tdra}
t_{H}\frac{d\ln r^{*}_{A}}{dt}=1.5[\cos(b\ln a)-1].
\end{equation}We notice that from the FIG.2, the AH displays an oscillating
while ascending trend, and there is remarkable ascending trend
without periodic variation for any period in $\omega$.
Furthermore, the ascending trends are similar at different epochs.
Namely, our Universe is not particular in the cosmological
evolution. FIG.3 shows that the AH does not change significantly
over one Hubble scale, so the equilibrium thermodynamics still can
be applied here.
\begin{figure}[!t]
\begin{center}
\includegraphics[width=8.5cm]{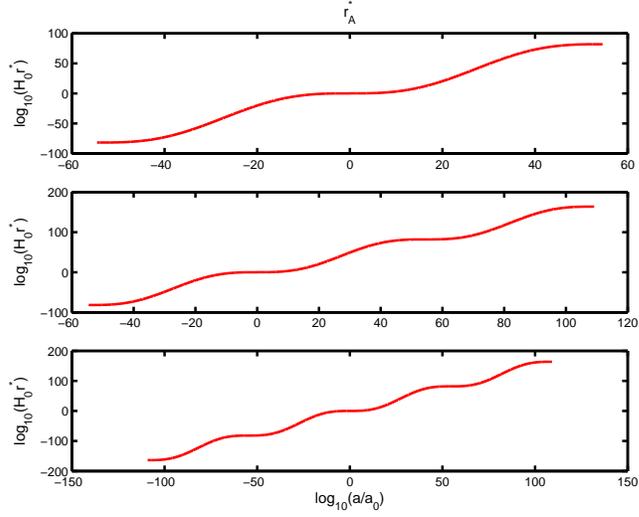}\\
\caption{The radius of an AH in units of $H_{0}^{-1}$ as function
of $a$ at different epochs, corresponding to three different
$\omega$ in FIG.1 respectively.}
\end{center}
\label{fig2}
\end{figure}

\begin{figure}[!t]
\begin{center}
\includegraphics[width=8.5cm]{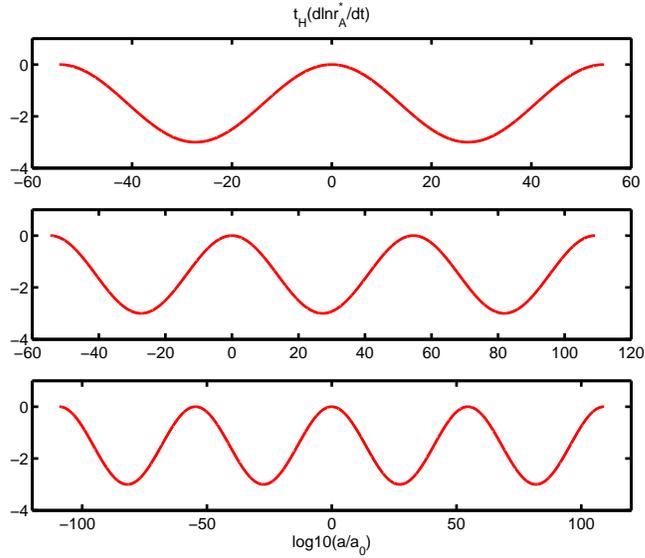}\\
\caption{The change of the radius of an AH $t_{H}d\ln
r^{*}_{A}/dt$ varies with time during a Hubble time at different
epochs, corresponding to three different $\omega$ in FIG.1
respectively.}
\end{center}
\label{fig3}
\end{figure}

\section{The Unified First Law of thermodynamics on the apparent horizon}
As a thermodynamical system, the Unified First Law of
thermodynamics $dE=TdS-PdV$ on the AH of the Undulant Universe
should be satisfied \cite{37}. Especially, the work term $PdV$
should be zero at infinity \cite{38}.

Now we calculate the flux of energy through the proper area of a
2-spheres $x^{0}=t$, $x^{1}=r^{*}$, where the time $t$ is allowed
to vary slowly, so, $t$ can be treated as constant. The surface
gravity on the AH is
\begin{equation}
\kappa =-f^{'}/2\mid _{r=r^{*}_{A}}=1/r^{*}_{A},  \label{k}
\end{equation}so the Hawking temperature is
\begin{equation}
T_{A}=\kappa /2\pi =1/2\pi r^{*}_{A}.\label{m6}
\end{equation}The amount of energy flux \cite{39} crossing the AH within the time
interval $dt$ is
\begin{eqnarray}
dE_{A} &=& 4\pi r^{*2}_{A}T^{\mu\nu}u_{\mu}u_{\nu}dt = 4\pi r^{*2}_{A}(\rho +P)dt\nonumber \\
\nonumber&=&1.5H_{0}^{-1}a^{0.5}[1-\cos(b\ln a)]\nonumber\\
&&\times\exp[-\frac{1.5}{b}\sin(b\ln a)]da.\label{de}
\end{eqnarray}The entropy on the AH is
\begin{equation}
  S_{A}=\mathcal{A}/4 = \pi r^{*2}_{A},\label{sa}
\end{equation}and its differential form is
\begin{equation}
dS_{A}=2\pi r^{*}_{A}dr^{*}_{A}.\label{dsa}
\end{equation}Thus we can obtain
\begin{eqnarray}
T_{A}dS_{A} &=&dr^{*}_{A}\nonumber \\
\nonumber &=&1.5H_{0}^{-1}a^{0.5}[1-\cos(b\ln a)]\nonumber \\
&&\times \exp[-\frac{1.5}{b}\sin(b\ln a)]da. \label{tdsa}
\end{eqnarray}Comparing Eq.(21) with Eq.(24), we have proved the result
$dE_{A}=T_{A}dS_{A}$. So the Unified First Law of thermodynamics
on the AH is confirmed, and the work that has been done is really
zero. From the FIG.4, we find no periodic variation in any periods
of $\omega$. FIG.5 demonstrates some details in the top panel of
FIG.4. At different epochs, as shown in FIG.5, the basic feature
for the change of the energy is similar, and $T_{A}dS_{A}$ and
$-dE_{A}$ are symmetrically distributed around the level line.
Actually, due to the oscillation of $\omega$, the variations of
energy are some fluctuations in the Standard Universe, which is
$\Lambda$CDM with $\Omega_{\Lambda0}=1$ and $\Omega_{M0}=0$.

However, if we add the matter term, or $k \neq 0$ to Eq(8), the
Unified First Law is broken. Namely, when the Universe is
deceleration, or the curvature becomes large generally near the
singularity, the Unified First Law needs more investigation. In
other words, for the Undulant Universe, the standard calculation
for the Unified First Law of thermodynamics is only confirmed in a
particular situation. We can speculate that the DE has scalar
vector form near the singularity \cite{44}. Another way is to find
some new calculation.

\begin{figure}[!t]
\begin{center}
\includegraphics[width=8.5cm]{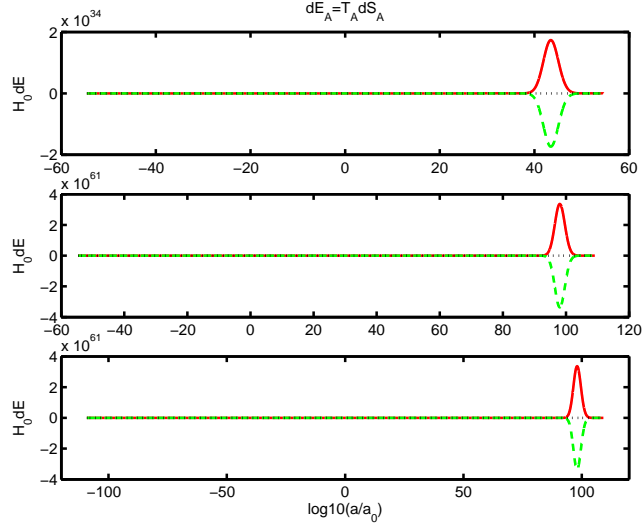}\\
\caption{$T_{A}dS_{A}$ in units of $H_{0}^{-1}$ (thick red line)
and $-dE_{A}$ in units of $H_{0}^{-1}$ (dashed green line) as
function of $a$ at different epochs, corresponding to three
different $\omega$ in FIG.1 respectively. And the level line is
the change of energy in the Standard Universe (dotted black
line).}
\end{center}
\label{fig4}
\end{figure}

\begin{figure}[!t]
\begin{center}
\includegraphics[width=8.5cm]{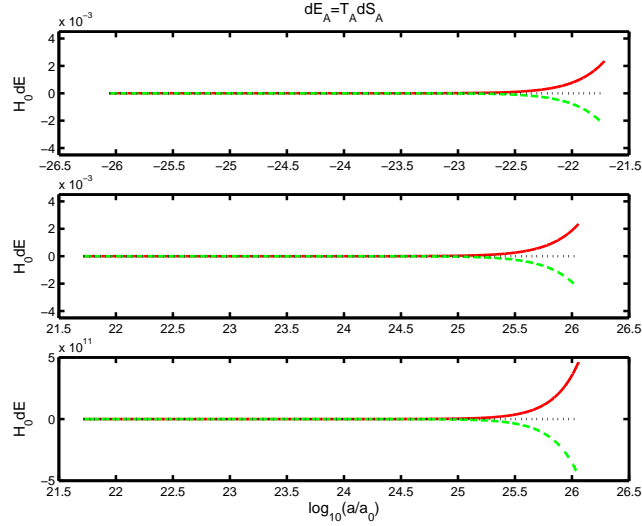}\\
\caption{$T_{A}dS_{A}$ (thick red line) in units of $H_{0}^{-1}$
and $-dE_{A}$ in units of $H_{0}^{-1}$ (dashed green line) as
function of $a$ at different epochs, which demonstrate some
details in the top panel of FIG.4. And the level line is the
change of energy in the Standard Universe (dotted black line).}
\end{center}
\label{fig5}
\end{figure}

\section{The Generalized Second Law of thermodynamics on the apparent horizon}
The preferred screen is the cosmological horizon. The AH is a good
cosmological horizon, and the Unified First Law of thermodynamics
on the AH is satisfied. So we can continue to discuss the
Generalized Second Law of thermodynamics.

For the Generalized Second Law, we now state a version of the
cosmic holographic principle based on the cosmological AH: the
particle entropy inside the AH can never exceed the AH
gravitational entropy, i.e. the entropy inside the AH plus the AH
gravitational entropy never deceases \cite{40}. We can obtain the
entropy of the Universe inside the AH through the Unified First
Law of thermodynamics:
\begin{eqnarray}
TdS_{i} &=& dE_{i}+PdV = Vd\rho+(\rho+P)dV.\label{ei}
\end{eqnarray}In terms of Eq.(25), the energy inside the AH is $E_{i} = 4\pi
r^{*3}_{A}\rho/3$, and the volume is $V = 4\pi r^{*3}_{A}/3$.
Combining Eq.(6) with Eq.(7), when $\Omega_{\Lambda0} = 1$, and
differentiating $\rho = \rho_{\Lambda}$, we get $d\rho =
3HdH/4\pi$. According to the Zeroth Law of thermodynamics
\cite{41}, we know that on a stationary surface in the
thermodynamical equilibrium, the temperature is constant Hawking
temperature. However, the temperature of the viscous matter is
higher than the Hawking temperature. At the same time, because the
Universe expands and the DE is dominant, the temperature declines
rapidly and becomes lower than the Hawking temperature. We might
define $T = uT_{H}$ \cite{42}, where $u$ is a real constant, $0 <
u \leq 1$. Indeed the parameter $u$ shows the deviation from
Hawking temperature. Therefore, we obtain
\begin{eqnarray}
  dS_{i} &=& \frac{Vd\rho+(\rho+P)dV}{T}= \frac{Vd\rho+(1+\omega)\rho dV}{uT_{H}}\\
  \nonumber &=& \frac{1.5H_{0}^{-1}}{u}\pi
a^{2}[3\cos(b\ln a)-1][\cos(b\ln a)-1] \nonumber \\
&&\times \exp[-\frac{3}{b}\sin(b\ln a)]da.\label{dsi}
\end{eqnarray}When $u=1$, $dS_{i}$ is minimum. So it is enough to study how the
minimal differential entropy evolves. For the entropy of the AH,
we get
\begin{eqnarray}
  dS_{A} &=& \frac{dr^{*}_{A}}{T_{A}}=3H_{0}^{-1}\pi a^{2}[1-\cos(b\ln a)]\nonumber \\
&&\times \exp[-\frac{3}{b}\sin(b\ln a)]da.\label{dsa}
\end{eqnarray}The total differential entropy is $dS_{total}=dS_{A}+dS_{i}$, and all of
the variations of the entropy vary with $a$. As shown in FIG.6,
$dS_{i} \leq 0$ all the time, but $dS_{A} \geq 0$ and $dS_{total}
\geq 0$. All of these cases at different epochs are similar. FIG.7
demonstrates some details in the top panel of FIG.6. we find
$dS_{i} \geq 0$ at different epochs in the small ranges. $dS_{i}$
are small fluctuations in the cosmological evolution.

Therefore, $dS_{total} \geq 0$ all the time, which is the minimal
total differential entropy. Correspondingly, the real total
differential entropy should not decrease. In a word, the
Generalized Second Law of thermodynamics is confirmed on the AH.
We know from \cite{44,45}, $\omega(a) < -1$ is meaningless
physically because of the negative entropy, while
$\omega(a)=-\cos(b\ln a) > -1$ all the time. The confirmation of
the Generalized Second Law for the Undulant Universe is another
support to this conclusion.

\begin{figure}[!t]
\begin{center}
\includegraphics[width=8.5cm]{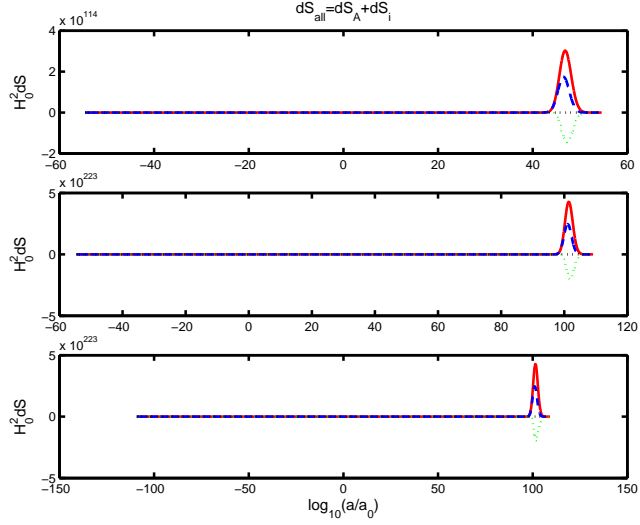}\\
\caption{$dS_{i}$ (thick dotted green line), $dS_{A}$ (thick solid
red line), and $dS_{i}+dS_{A}=dS_{total}$ (thick dashed blue line)
in units of $H_{0}^{-2}$ as functions of $a$ at different epochs,
corresponding to three different $\omega$ in FIG.1 respectively.
The varied entropy of the Standard Universe is the level line
(thin dotted black line).}
\end{center}
\label{fig6}
\end{figure}

\begin{figure}[!t]
\begin{center}
\includegraphics[width=8.5cm]{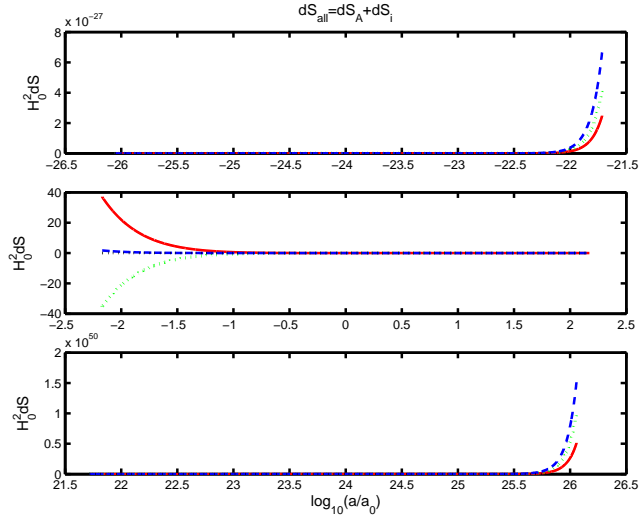}\\
\caption{$dS_{i}$ (thick dotted green line), $dS_{A}$ (thick solid
red line), and $dS_{i}+dS_{A}=dS_{total}$ (thick dashed blue line)
in units of $H_{0}^{-2}$ as functions of $a$ within three
different small ranges, and at different epochs, which demonstrate
some details in the top panel of FIG.6. The varied entropy of the
Standard Universe is the level line (thin dotted black line).}
\end{center}
\label{fig7}
\end{figure}

\section{Conclusion}
The properties of DE have been studied since long time ago, and
many models have been put forward. Although there is no evidence
to give an exact model consistent with the real world, researchers
have been successful in many aspects. In this paper, we study the
Undulant Universe on its thermodynamical properties.

We have shown that the Undulant Universe is a model to solve the
cosmic coincidence problem without fine tuning, i.e. nowadays
phase of the Universe is not particular. Thus we can come from any
phase in the past, and go to any phase in the future. We find that
neither the EH nor the PH exists in the Undulant Universe, which
is consistent with our conclusion above. However, we find another
cosmological horizon, the AH, with a definition coming from
topology. We choose a special expression for the AH, whose
evolution can be seen from FIG.2. There is no periodicity. No
matter how the Universe evolves, we can always find a AH, and all
of their evolution is similar. So we can consider the Universe as
a thermodynamical system inside the AH. And because of no
significant change over one Hubble scale, as shown in FIG.3, the
equilibrium thermodynamics can be used.

The Four Laws of BH thermodynamics \cite{41} are successful.
However these conjectures should be proved in the Universe. The
thermodynamical First Law $dM=TdS$ on the AH is expressed as
$dE_{A}=T_{A}dS_{A}$. Because the work term $PdV$ is zero at
infinity, we have ignored it. According to our deduction, the
Unified First Law of thermodynamics is satisfied perfectly. As
shown in FIG.4, there is no periodic variation in any periods of
$\omega$, and the basic feature for the change of the energy is
similar at different epochs. FIG.5 presents the variations of
energy in small ranges, and there are some fluctuations comparing
with the Standard Universe. The fluctuations are due to the
oscillation of $\omega$.

Holography can be applied to the standard cosmological context, so
we consider the world as a hologram. The entire information about
the global 3+1-dimensional spacetime can be stored on particular
hypersurfaces. The connection between entropy and information is
well known \cite{43}. Accordingly, the total entropy of the
Universe increases in the process. We obtain the entropy inside
the AH by the Unified First Law of thermodynamics. Because the
temperature of the Universe dominated by the DE is not higher than
the Hawking temperature, we choose the maximal temperature to get
the minimal total differential entropy. The total differential
entropy is calculated from $dS_{total}=dS_{A}+dS_{i}$. The
temperature of the cosmological material will decline as the
Universe expands, so $dS_{i} \leq 0$. Its accompanying loss of
energy can cause a back reaction on the cosmic dynamics, so that
the horizon area increases, and $dS_{A} \geq 0$. We find that the
total entropy of the AH does not decrease with time, $dS_{total}
\geq 0$. In the small ranges, the entropy of the visible Universe
decreases at different epochs, $dS_{i} \geq 0$, which locally are
small fluctuations in the cosmological evolution. Thus the
Generalized Second Law of thermodynamics is satisfied on the AH.
As shown in FIG.6 and FIG.7, whenever and wherever the Universe
evolves, the variations of the entropy are similar. Although we
just discuss the minimal total differential entropy, the
Generalized Second Law of thermodynamics remains confirmed in the
condition of the real world.

In conclusion, the AH of the Undulant Universe is a good
holographic screen, and can be seen as a boundary of keeping
thermodynamical properties. The Undulant Universe is in thermal
equilibrium inside the AH, and behaves very well in its
cosmological evolution. In spite of some undulant departure from
the Standard Universe, the Undulant model solves some coincidence
problems in the Standard Universe.

In summary, the investigation into the Universe is very intriguing
from the thermodynamical perspective, however, there are still
some problems to be solved. In fact, the acceleration of our
Universe is temporary, and there ever was and will be a matter
dominated phase or singularity. Given this we should consider
matter dominated or large curvature case. As we know, the
conversation law always works. When adding the matter term and $k
\neq 0$ to the Undulant Universe, e.g., when the Universe is
decelerating or near singularity, the law is broken. This bad
result means that, the energy in the Undulant Universe
$\omega(a)=-\cos(b\ln a)$ maybe some other form at these epochs to
keep the conversation law confirmed. Another possibility is that
the entropy should be redefined. However, the success of our
standard deduction encourages the definition in the Undulant
Universe. Furthermore, thermodynamical properties demand to be
explored more deeply in the future, and there are some more
cosmological horizons expected to be studied.

\section{acknowledgements}
Tian Lan would like to thank Prof. Canbin Liang for his patient
and valuable teaching, and Yongping Zhang for her discussion and
suggestions, and all the students in her lab for their valuable
comments and help. This work was supported by the National Science
Foundation of China (Grants No.10473002), the Ministry of Science
and Technology National Basic Science program (project 973) under
grant No.2009CB24901 and the Scientific Research Foundation for
the Returned Overseas Chinese Scholars, State Education Ministry.

\end{document}